\documentclass[10pt]{article}
\usepackage{amsmath,amssymb}
\usepackage{graphicx}

\newcommand{\Id}{\mathbf 1} 
\newcommand{\e}{\boldsymbol e} 
\newcommand{\op}[1]{\boldsymbol{#1}} 
\newcommand{\sgm}{\boldsymbol\sigma}

\newcommand{\Eq}[1]{Eq.~(\ref{#1})}
\newcommand{\Fig}[1]{Fig.~\ref{Fig:#1}}

\title{\bf Hyperspheres and control of spin chains}
\author{\em Alexander Yu.\ Vlasov\\
Federal Radiological Center (IRH),
197101 Mira Street 8, St.--Petersburg, Russia}
\date{6 December 2004}
\begin{document}
\sloppy
\twocolumn[
\maketitle
\centering
{\bf\large Abstract}
\begin{quote}
 Here is considered application of Spin$(m)$ groups in theory of quantum 
control of chain with spin-$\frac{1}{2}$ systems. It may be also compared
with $m$-dimensional analogues of Bloch sphere, but has nontrivial distinctions
for chain with more than one spin system.
\end{quote}
\bigskip
]

\section{Introduction}
It is convenient to use sphere to represent state of one spin-$1/2$ system.
It is so called Bloch or Poincar\'e sphere. The state of space of $n$ spin-$1/2$ systems
is complex Hilbert space with dimension $2^n$, but due to normalization condition
there are $2^n-1$ complex or $2\,(2^n-1)$ real parameters. So for $n=1$ there
are $2=2(2^1-1)$ real parameters, {\em e.g.} two Euler angles describing point
on surface of sphere. Any unitary transformation of spin-$1/2$ system
corresponds to rotation of Bloch sphere in agreement with $2 \to 1$ isomorphism
of groups SU$(2)$ and SO$(3)$.

\begin{figure}[htb]
\includegraphics[scale=0.5]{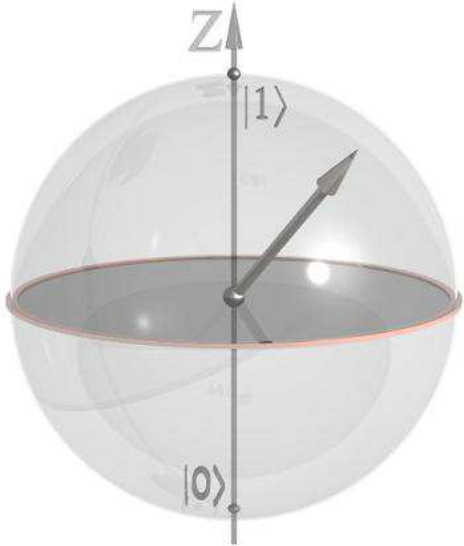}
\caption{Bloch sphere.}
\label{Fig:bloch}
\end{figure}

For $n>1$ there is no such convenient visualization of space of {\em all} states 
with higher dimensional spheres\footnote{But see {\bf Note} on page \pageref{Note}.}, 
but exist some interesting subspaces with such
property. The subspaces are important for theory of quantum computations and
control, because for some physical systems they may correspond to 
{\em simpler accessible} set of physical states. 

\section{Rotations and Spin groups}

\begin{figure*}[htb]
\includegraphics[scale=0.75]{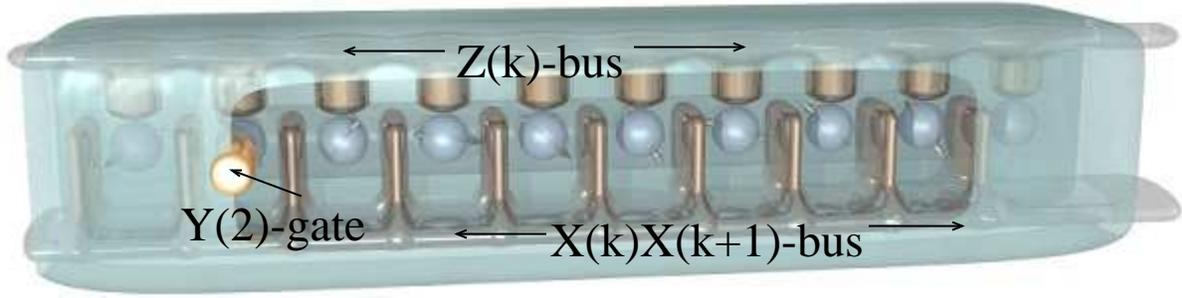}
\caption{Scheme of spin chain with control buses.}
\label{Fig:spins}
\end{figure*}

Let us consider for chain with $n$ qubits set with $2n$ Hermitian matrices 
\begin{equation}
\begin{split}
 \e_{2k} & = 
  {\underbrace{\sgm_z\otimes\cdots\otimes \sgm_z}_k\,}\otimes
 \sgm_x\otimes\underbrace{\Id\otimes\cdots\otimes\Id}_{n-k-1} \, ,
 \\
 \e_{2k+1} & = 
 {\underbrace{\sgm_z\otimes\cdots\otimes \sgm_z}_k\,}\otimes
 \sgm_y\otimes\underbrace{\Id\otimes\cdots\otimes\Id}_{n-k-1} \, .
 \label{defE}
\end{split}
\end{equation}

The set is well known in quantum mechanics due to Jordan, Wigner and Weyl works 
\cite{WeylGQM}, because operators
\begin{equation}
\op a_k = \frac{\e_{2k}+i\e_{2k+1}}{2},
\quad
\op a_k^\dag = \frac{\e_{2k}-i\e_{2k+1}}{2}
\label{aferm}
\end{equation}
({\em i.e.}, $2^n \times 2^n$ complex matrices)
provide representation of {\em canonical anticommuting relations} (CAR)
\begin{equation}
\{\op a_k,\op a_j\}=\{\op a_k^\dag,\op a_j^\dag\}=0,
\quad
\{\op a_k,\op a_j^\dag\} = \delta_{kj}.
\label{CAR}
\end{equation}

On the other hand, \Eq{defE} may be used for construction Spin$(2n)$
and Spin$(2n+1)$ groups \cite{ClDir}. The Spin$(2n+1)$ groups has 
$2 \to 1$ isomorphism with group of rotations SO$(2n+1)$ and so
for $n=1$ we have usual model with Bloch sphere and group SO$(3)$ of
three-dimensional rotations of the sphere.

For $n>1$ it is also possible to consider groups Spin$(2n+1)$ and 
SO$(2n+1)$, but they may not describe all possible transformations 
of system with $n$ qubits. Such transformations may be 
described by huge group SU$(2^n)$ with dimension $4^n - 1$, but
group SO$(2n+1)$ has dimension $(2n+1)n$ and only for $n=1$ both
numbers coinside:
$$
\begin{array}{||c||r|r|r|r|r|r||}
\hline
     n& 1 &  2  & 3 & 4 & 5 & 10 \\[-\doublerulesep]
\hline\hline
(2n+1)n & 3 & 10 & 21 & 36 & 55 & 210 \\
\hline
4^n - 1 & 3 & 15 & 63 & 255 & 1023 & 1048575\\
\hline
\end{array}
$$

The \Eq{defE} are Hermitian matrices and may be considered as set of $2n$
Hamiltonians for system with $n$ qubits. If to use {\em only} these 
Hamiltonians for control of system, then unitary evolution belong only 
some subgroup of $S_o \subset {\rm SU}(2^n)$, {\em i.e.}, the control is 
{\em not universal}. Despite the subgroup $S_o$ belongs to such exponentially
big space, it is ismorphic to Spin$(2n+1)$ \cite{Clif} and due to usual 
relation of Spin groups with rotations of $2n+1$-dimensional hypersphere
may be considered as higher dimensional analogue model of Bloch sphere 
rotations.

It should be mentioned, that together with $S_o \subset {\rm SU}(2^n)$,
$S_o \cong {\rm Spin}(2n+1)$, it is also useful to consider 
(maybe more familiar) {\em even} subgroup $S_e \subset S_o$,
$S_e \cong {\rm Spin}(2n)$ \cite{ClDir,Clif}. The subgroup is generated by
even elements $\op d_k = i\e_k \e_{k+1}$
\begin{subequations} \label{defD}
\begin{align}
   \op{d}_{2k} &= 
  {\underbrace{\Id\otimes\cdots\otimes \Id}_k\,}\otimes
 \sgm_z\otimes\underbrace{\Id\otimes\cdots\otimes \Id}_{n-k-1}\, , 
 \label{defD1}
 \\
\!\!\!
  \op{d}_{2k+1} &= 
  {\underbrace{\Id\otimes\cdots\otimes \Id}_k\,}\otimes
 \sgm_x\otimes\sgm_x\otimes\underbrace{\Id\otimes\cdots\otimes \Id}_{n-k-2} . 
 \label{defD2}
\end{align}
\end{subequations}

Now it is possible to add any operator $\e_k$, say
\begin{equation}
 \e_0 = \sgm_x\otimes\underbrace{\Id\otimes\cdots\otimes\Id}_{n-1}\, , 
 \label{E0}
\end{equation} 
to provide control on $S_o$ and it was shown in \cite{Clif},
that it is enough to add also any {\em third (or fourth) order} operator 
like $\e_k\e_l\e_m$, say
\begin{equation}
 \e_0\e_1\e_3 =  
 \Id\otimes\sgm_y\otimes\underbrace{\Id\otimes\cdots\otimes\Id}_{n-2}\, , 
 \label{E013}
\end{equation} 
to provide universal control, whole group SU$(2^n)$.

The subgroup $S_e$ also may be generated by Hermitian bilinear 
combinations of fermionic annihilation and creation operators 
\Eq{aferm}, {\em i.e.}
\begin{equation}
 \op a_j \op a_k^\dag +  \op a_k \op a_j^\dag,
\quad
\op a_j \op a_k +  \op a_k^\dag \op a_j^\dag.
\label{bferm}
\end{equation}

It should be mentioned, that \Eq{aferm} and \Eq{bferm} here should be
considered rather from point of view of {\em simulation} of quantum control
and computations with fermionic systems \cite{VlaTMR,BK00}, because most 
methods used above may be applied to arbitrary system of $n$ qubits and are 
not related directly with fermionic statistic of particles in spin chain. 

The note about simulation may be quite essential, say in usual 
fermionic systems \cite{BK00} as well as in {\em linear optics} 
KLM \cite{KLM} model of quantum computing (close related with the
fermionic operators \cite{TD01}) appearance of the group $S_e$ generated
by bilinear combinations produces specific difficulty. Really, universal
control suggest exponentially big space of parameters ($4^n$), but
group $S_e$ has dimension only quadratic with respect to number of 
systems ($\dim S_e = 2n^2-n < \dim S_o = 2n^2+n 
\ll \dim {\rm SU}(2^n) = 4^n-1$). 

For spin chain considered here the problem with non-universality is not 
so essential, because it is enough to use Hamiltonians \Eq{E0} and
\Eq{E013} to extend group $S_e$ to exponentially big group
of universal control. And the two extra Hamiltonian are simply
one-qubit rotations $\sgm_x$ and $\sgm_y$ with first and second qubit and so
complexity of realization for such operations may not exceed analogous
operations $\op d_{2k}$ with Hamiltonian $\sgm_z$ \Eq{defD1}.

On the other hand, in fermionic computations and linear optics,
analogues of operators with higher order like \Eq{E013} may not
be realized so simply because need physical processes with very
law amplitude, like nonlinear $\gamma+\gamma$ interactions. 
So spin chain not only may simulate some fermionic or optic
computations and control available for modern state of technologies, 
but also some currently inaccessible, very weak processes.

Using modern jargon whole universal set of gates considered above
may be denoted
\begin{equation}
\begin{array}{rll}
{\bf I}) \quad &\op Z(k+1) &= \op{d}_{2k}, \\
{\bf I\!I})\quad &\op X(k+1)\op X(k+2) &= \op{d}_{2k+1}\\
         &\op X(1) &= \e_0\\
{\bf I\!I\!I}) \quad &\op Y(2) &= \e_0\e_1\e_3,
\end{array}
\label{III}
\end{equation}
(see \Fig{spins}),
where gate $\op X(1)$ for simplicity considered as part of
$\op X(k+1)\op X(k+2)$-bus. In such a case buses {\bf I} and {\bf I\!I}
with $2n$ gates generate subgroup $S_o$ and may be associated
with control of rotations in dimension $D=2n+1$.

Such method let us not only model some processes in linear optics
and fermionic systems, but also decomposes difficult task of
control with exponentially big group SU$(2^n)$ on simpler
task of control with groups SO$(2n)$ or SO$(2n+1)$ and SO$(2)$.

Let us consider structure of group $S_o$ with more details. Unitary
matrices from groups $S_o$ and $S_e$ may be represented as
\begin{equation}
\Bigl\{U \in S_e \Big| 
U = \exp\bigl(\sum_{j=0}^{2n-1}\sum_{k=0}^{j-1}\! b_{kj} \op e_k \op e_j\bigr) \Bigr\}, 
\end{equation}
\begin{equation}
\Bigl\{U \in S_o \Big|
U = \exp\bigl(i \! \sum_{j=0}^{2n-1}\! b_j \op e_j + \!\!
\sum_{\substack{j,k=0\\k<j}}^{2n-1}\! b_{kj} \op e_k \op e_j\bigr)  \Bigr\}. 
\end{equation}
The number of parameters is in agreement with dimensions mentioned
above $\dim S_e = 2n^2-n$, $\dim S_o = 2n^2+n$. 

On the other hand, due to general defininion of group for any 
$U_1, U_2 \in S_o$ product $U_1 U_2 \in S_o$, but 
$i \e_k = \exp(i \frac{\pi}{2} \e_k) \in S_o$ and so $2n$ elements $\e_k$ 
and all $2^{2n}$ possible products of the elements (up to 
insignificant multiplier $i$) belong to $S_o$.

The $2^{2n}$ products of $\e_k$ are simply $4^n$ possible {\em tensor 
products} with $n$ terms. Each term is either Pauli martix or 
$2 \times 2$ unit matrix. It is discrete {\em Pauli group}, widely used 
in theory of quantum computations \cite{GotHeis}. 

So the discrete Pauli group with $4^n$ elements is subgroup of continuous
group $S_o$. The property show, that $S_o$ has rather nontrivial structure,
because matrices from Pauli group are {\em basis} in $4^n$ dimensional space
of matrices and so {\em convex hull} of $S_o$ is also $4^n$-dimensional.


Anyway, $S_o$ may be described as subspace of SU$(2^n)$ isomorphic
to SO$(2n+1)$ and so quantum control with exponentially big 
space of parameters may be considered as alternating of
control over ``winding'' subspace $S_o$ isomorphic to SO$(2n+1)$
and one-parametric rotations $e^{i \alpha \op Y(2)}$.   
The control over $S_o$ represented on \Fig{spins} by two buses and
one-parametric rotations as one ``exceptional'' gate.

\paragraph{Note (June 2005)}\label{Note} It should be mentioned 
a specific case, corresponding to two qubits and not discussed
in presented paper. Due to `sporadic' isomorphism 
$${\rm SU}(4) \cong {\rm Spin}(6)$$
transformations of two qubits may be associated with rotations of 6D sphere
and so it is also relevant to considered theme. 
It may be described using so-called {\em Klein correspondence} and
{\em Pl\"ucker coordinates} (introduced first in few papers written between 1865 and 1870 by 
F. Klein and J. Pl\"ucker, see R.~Penrose and W. Rindler, {\em Spinors and Space-Time,
{\bf vol.2}, Spinor and Twistor Methods in Space-Time Geometry}, Cambridge Univ.\ Press 1986),
but it should be discussed elsewhere.

\end{document}